\documentclass[aps,twocolumn]{revtex4}
\usepackage{graphicx}
\usepackage{amsmath,amsfonts}
\begin{document}

\title{\large Numerical indications on the semiclassical limit of the flipped vertex\\[0.4cm]}
\author{Elena Magliaro{\it ${}^{ab}$}, Claudio Perini{\it ${}^{ac}$}, Carlo Rovelli{\it ${}^a$}\vspace{0.2cm}}
\affiliation{\small\it ${}^a$Centre de Physique Th\'eorique de Luminy\footnote{Unit\'e mixte de recherche (UMR 6207) du CNRS et des Universit\'es
de Provence (Aix-Marseille I), de la M\'editerran\'ee (Aix-Marseille II) et du Sud (Toulon-Var); laboratoire affili\'e \`a la FRUMAM (FR 2291).}, Case 907, F-13288 Marseille, EU\\
\small\it ${}^b$Dipartimento di Fisica, Universit\`a degli Studi Roma Tre, I-00146 Roma, EU\\
\small\it ${}^c$Dipartimento di Matematica, Universit\`a degli Studi Roma Tre, I-00146 Roma, EU}

\date{\vspace{0.25cm}\small\today}

\begin{abstract}\noindent
We introduce a technique for testing the semiclassical limit of a quantum
gravity vertex amplitude. The technique is based on the propagation of a
semiclassical wave packet.  We apply this technique to the newly introduced 
``flipped" vertex in loop quantum gravity, in order to test the intertwiner 
dependence of the vertex.  Under some drastic simplifications, we find 
very preliminary, but surprisingly good numerical evidence for the correct 
classical limit.\\[0.1cm]
\end{abstract}

\maketitle
\noindent

Suppose you are explicitly given the propagation kernel $W_t(x,y)$ of a
one-dimensional nonrelativistic quantum system defined by a hamiltonian
operator $H$
\begin{equation}
    W_t(x,y)=\langle x|e^{-\frac \imath\hbar Ht} |y\rangle
\end{equation}
and you want to study whether the classical ($\hbar\to 0$) limit of this
quantum theory yields a certain given classical evolution.  One of 
the (many) ways of doing so is to propagate a wave packet $\psi_{x,p}(x)$ with 
$W_t(x,y)$.  Suppose that in the time interval $t$ the classical theory evolves  
the initial position and momentum $x_i, p_i$ to the final values  $x_f, p_f$.  
Then you can consider a semiclassical wave packet  $\psi_{x_i,p_i}(y)$  
centered on the initial values $x_i, p_i$, compute its evolution under the 
kernel
\begin{equation}
    \phi(x):=\int dy\  W_t(x,y)\ \psi_{x_i,p_i}(y)   
\end{equation}
and ask whether or not this state is a semiclassical wave packet  
centered on the 
correct final values $x_f, p_f$.  In this letter, we consider the possibility of
using this method for exploring the semiclassical limit of the dynamics of
nonperturbative quantum gravity. 

An explicit expression for the vertex amplitude in loop quantum gravity
has been recently proposed in \cite{EPR}; we call it here the ``flipped"
vertex, following  \cite{EPR2}. This expression corrects certain difficulties that 
have emerged with the Barrett-Crane vertex amplitude \cite{I}.   In particular, the 
Barrett-Crane vertex was shown to have a good semiclassical
behavior as far as its dependence on \emph{spin} variables was concerned, but
not in its \emph{intertwiner} sector.  The flipped vertex $W(j_{nm},i_n)$ is a 
function of ten spin variables $j_{nm}$ where $n,m=1,...,5$ and five intertwiner variables
$i_n$ and it is hoped to correct the intertwiner dependence of the Barrett-Crane vertex.  
Here we investigate this intertwiner dependence.  A number of variants of the 
flipped vertex have appeared in literature \cite{LS}, 
but we do not consider these variants here. 

The derivation of the vertex amplitude presented in \cite{EPR2} indicates that 
the process described by one vertex can be seen as the dynamics of a single cell 
in a Regge triangulation of general relativity.   This is a fortunate situation, because it
allows us to give a simple and direct geometrical interpretation to the dynamical 
variables entering the vertex amplitude, and a simple formulation of the 
dynamical equations.   The boundary of a Regge cell is formed by five tetrahedra
joined along all their faces, thus forming a closed space with the topology of a 
3-sphere.  Denote $A_{nm}$ the area of the triangle $(nm)$ that separates 
the tetrahedra $n$ and $m$.  Denote $\alpha_{n}^{(mp,qr)}$ the angle between 
the triangles $(mp)$ and $(qr)$ in the tetrahedron $n$.  And denote $\Theta_{nm}$
the angle between the normals to the tetrahedra $n$ and $m$.   These quantities 
determine entirely the intrinsic ($A_{nm},\alpha_{n}^{(mp,qr)}$) and extrinsic 
($\Theta_{nm}$) classical geometry of the boundary surface. 

The ten spins $j_{nm}$ are the quantum numbers of the areas $A_{nm}$ \cite{EPR2,lqg2}.  
The five intertwiners $i_{n}$ are the quantum 
numbers associated to the angles $\alpha_{n}^{(mp,qr)}$. More precisely, they are the
eigenvalues of the quantity
\begin{equation}
    i_n^{(mp,qr)} = A_{mp}^2+A_{qr}^2+2A_{mp}A_{qr}\cos\alpha_{n}^{(mp,qr)}.
\end{equation}
Each tetrahedron has six such angles, of which only two are independent
(at given values of the areas); but the two corresponding quantum operators 
do not commute \cite{tetraedro} and a basis of the Hilbert space on which they act can be 
obtained by diagonalizing just a single arbitrary one 
among these angles.  Therefore the intrinsic geometry of the boundary of a 
classical Regge cell is determined by twenty numbers, but the the corresponding
quantum numbers are only fifteen: the fifteen quantities $j_{nm},i_n$. These are the 
fifteen arguments of the  vertex.  
When using the intertwiners $i_n$, we have of course to specify to 
which pairing $ i_n^{(mp,qr)}$ we are referring. 

The equations of motion of any dynamical system can be expressed as 
constraints on the set formed by the initial, final and (if it is the case) 
boundary variables.  For instance, in the case of the evolution of a free 
particle in the time interval $t$, the equations of motion can be expressed 
as constraints on the set $(x_i, p_i, x_f, p_f)$. These constraints are of 
course $m(x_f-x_i)/t=p_i=p_f$. (For the general logic of this approach 
to dynamics, see \cite{lqg}.) In general relativity, the Einstein equations
can be seen as constraints on boundary variables $A_{nm},\alpha_{n}^{(mp,qr)}$
 and $\Theta_{nm}$. These, in fact, can be viewed as the ensemble of the initial,
boundary and final data for a process happening inside the boundary 
3-sphere.   Such constraints are a bit difficult to
write explicitly, but one solution is easy: the one that corresponds to 
flat space and to the boundary of a \emph{regular} 4-simplex.  This is given 
by all equal areas $A_{nm}=j_0$, all equal angles $i_n=i_0$, and 
$\Theta_{nm}=\Theta$, where elementary geometry gives
\begin{equation}
i_0=\frac{2}{\sqrt 3}\ j_0,\ \ \ \ \ 
\cos\Theta=-\frac{1}{4}. 
\end{equation}

It follows immediately from these considerations that a boundary wave packet centered 
on these values must be correctly propagated by the vertex amplitude, if the 
vertex amplitude is to give the Einstein equations in the classical limit. 

The simplest wave packet we may consider is a diagonal gaussian wave packet
\begin{equation}
\psi(j_{nm},i_n)=\prod_{nm}\Psi(j_{nm})\prod_{n}\psi(i_n)
\label{stato}
\end{equation}
where 
\begin{equation}
\Psi(j_{nm})= e^{-\frac{1}{\tau}(j_{nm}-j_0)^2+\imath\Theta j_{nm}}
\end{equation}
and
\begin{equation}
\psi(i)= N \sqrt{d_i}\;e^{-\frac{1}{\sigma}(i-i_0)^2+\imath \theta i}. 
\end{equation}
The normalization factors $\sqrt{d_i}=\sqrt{2i+1}$, here and below, are 
required by the conventions we use in this paper, where the 3j-symbols 
are normalized. For details on the conventions, see \cite{I}.
The constants $\sigma$ and $\theta$ are fixed by the requirement
that the state is peaked on the value $i_n=i_0$ also when we change the
pairing at the vertex. That is, by the requirement that \emph{all} angles of the
tetrahedron are equally peaked on $i_n=i_0$. It was shown in \cite{sc} that this
requirement fixes these constants to the values
\begin{equation}
\sigma=\frac{3}{4j_0}\ ,\ \ \ \ \ 
\theta=\frac{\pi}{2}. 
\end{equation}
In other words, the state considered is formed by a gaussian state on the spins,
with $\Theta$-phases given by the extrinsic curvature and by a ``coherent 
tetrahedron" state (see \cite{sc})
\begin{equation}
\psi(i)= N \sqrt{d_i}\;e^{-\frac{3}{4j_0}(i-i_0)^2+\imath \frac\pi 2 i} 
\label{coherent}
\end{equation}
for each tetrahedron.

Let us write the wave packet (\ref{stato}) as an ``initial state" times a ``final 
state" by viewing the process represented by the spacetime region described 
by the Regge cell as a process evolving four tetrahedra into one.  That is, let us write 
this state in the form 
\begin{equation}
\psi(j_{nm},i_n)=\psi_{init}(j_{nm},i'_n)\psi(i_5)
\end{equation}
where $i'_n=(i_1,...,i_4)$.   Then we can test the classical limit of the vertex amplitude 
by computing the evolution of the four ``incoming" tetrahedra generated by the
vertex amplitude
\begin{equation}
\phi(i)=\sum_{j_{nm},i'_n} W(j_{nm},i'_n, i) \psi_{init}(j_{nm},i'_n)
\end{equation}
where $i$ is $i_5$, and comparing $\phi(i)$ with $\psi(i)$. If the vertex amplitude has general 
relativity as its classical limit, then we expect that in the semiclassical limit, namely
for large $j_0$,  the evolution should 
evolve the ``initial" boundary state $\psi_i(j_{nm},i'_n)$ into a final state $\phi(i)$ 
which is still a wave packet centered on the same classical tetrahedron as the 
state $\psi(i)$ given in (\ref{coherent}). That is,  $\phi(i)$ must be a state ``similar" 
to  $\psi(i)$, plus perhaps quantum corrections representing the quantum spread 
of the wave packet. 

We have tested this hypothesis numerically, under a drastic approximation: replacing
the gaussian dependence on the spins with a state concentrated on $j_{nm}=j_0$. That
is, we have tested the hypothesis in the $\tau\to\infty$ limit.  We are not sure this
approximation makes sense, because in this limit the $\Theta$ phase dependence
drops, and this may jeopardize the semiclassical limit. This approximation is therefore 
only a first tentative hypothesis, that allows us to perform numerical calculations.
Numerical calculation have proven useful in quantum gravity \cite{numerical},
and the numerical investigation of the new vertex has already begun  \cite{numerical2};
here we consider a different angle from which the problem can be explored. 

The hypothesis we want to test is thus the following. We want to compare the evolved state
\begin{equation}
\phi(i)=\sum_{i_1... i_4} W(i_1,...,i_4, i) 
\prod_{n=1}^4 \psi(i_n)
\end{equation}
with the coherent tetrahedron state (\ref{coherent}), where 
\begin{equation}
W(i_n):=
W(j_{nm},i_n)|_{j_{nm}=j_0}.  
\end{equation}
If the function $\phi(i)$ turns out to be sufficiently close to the 
coherent tetrahedron state $\psi(i)$, we can say that, under the hypotheses given, 
the flipped vertex amplitude appears to evolve four coherent tetrahedra into one 
coherent tetrahedron, consistently with the flat solution of the classical Einstein 
equations. 

From \cite{EPR2} the flipped vertex reads in the present case
\begin{equation}
W(i_n) = \! \sum_{i_n^{\scriptscriptstyle +},i_n^{\scriptscriptstyle -}}
15j\!\left({\frac{j_{0}}{2}}, i^+_n\right)
15j\!\left({\frac{j_{0}}{2}}, i^-_n\right)\prod_n f^{i_n}_{i_n^{\scriptscriptstyle +}
i_n^{-}}\nonumber
\label{A}
\end{equation}
where 
\begin{equation}
f^{i}_{i^{\scriptscriptstyle +}i^{-}}\!=\!
\sqrt{d_id_{i^{\scriptscriptstyle +}}d_{i^{\scriptscriptstyle -}}}
\;i_{a_1...a_4} i^{\scriptscriptstyle +}_{b_1...b_4}
i^{-}_{c_1...c_4}\prod_{n=1}^4 \! \left(
\begin{array}{ccc}
j_0&\frac{j_0}2&\frac{j_0}2\\ a_n& b_n& c_n
\end{array}
\right)\!.  \nonumber
\end{equation}
The 15j-symbol is the contraction of five 4-valent intertwiners,
each constructed with two Wigner 3j-symbols and each
normalized with a $\sqrt{d_{i}}$ factor. The matrix 
indicates the Wigner 3j-symbol. The indices $a_n$ are 
in the representation $j_0$ and the indices $b_n$ and $c_n$ in the 
representation $j_0/2$.  
The quantities $i_{a_1...a_4} $ are the components of the
invariant tensors that define the intertwiner $i$. For technical details, see \cite{I}.

We have compared the two functions $\psi(i)$ (coherent tetrahedron) 
and $\phi(i)$ (evolved state) for the cases $j_0=2$
and $j_0=4$. The numerical results are shown in the figures below.
The overall relative amplitude of  $\psi(i)$ and $\phi(i)$ is freely adjusted
by fixing the normalization constant $N$ and therefore is not significant. The 
quantity $i_{\text{mean}}$ is the
mean value of $i$. It gives the position of the wave packet. 
The quantity $\sigma/2$ is the corresponding variance. It gives the
(half) width of the wave packet.   In Fig.1 and Fig.3 we compare the modulus 
square of the wave function (for the two values of $j_0$). In Fig.2 and Fig.4 
we compare the modulus square of the discrete Fourier transform of the wave 
function:   $n$ stands for the $n^{\text{th}}$ multiple of the fundamental frequency 
$2\pi/j_0$.

\noindent 
Case $j_0=2:$

\vspace{0.5cm}
\centerline{\includegraphics[height=3cm]{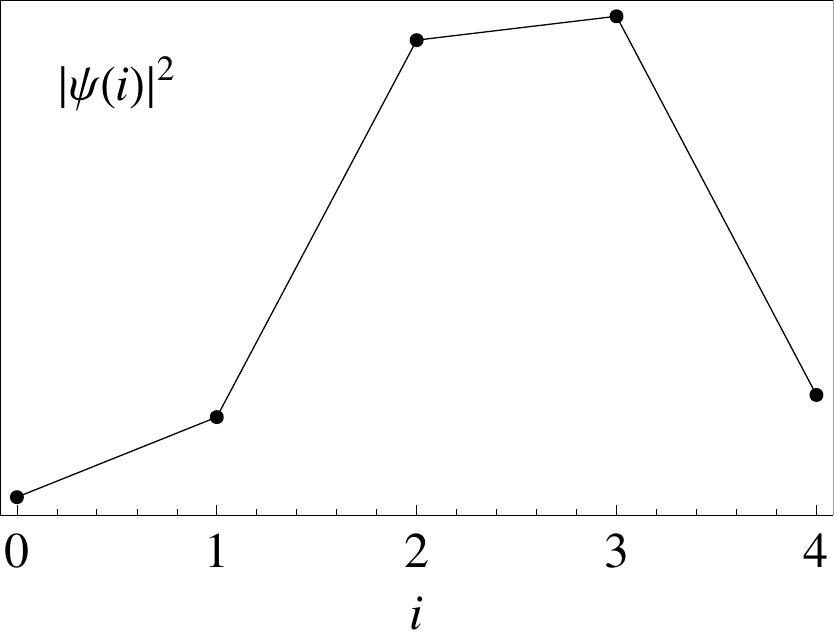}
\hspace{1em} \includegraphics[height=3cm]{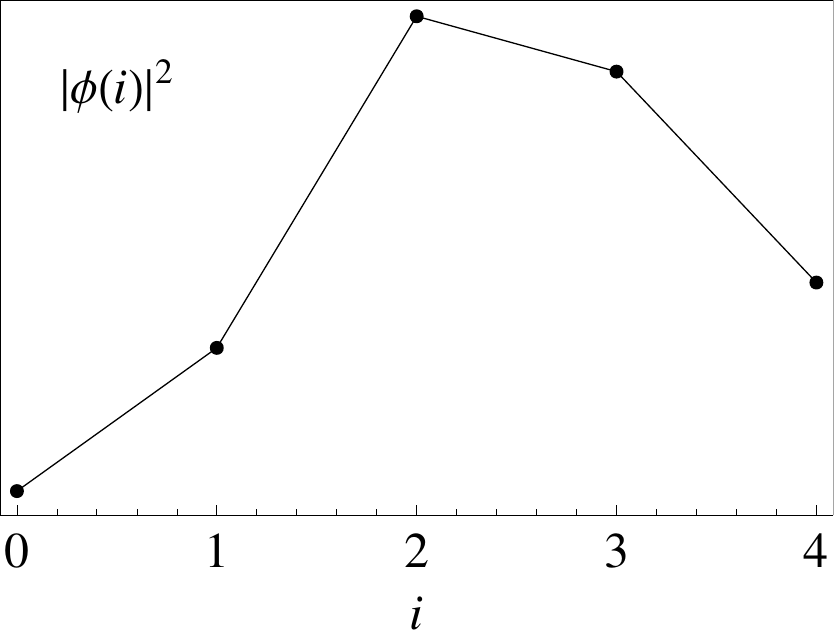}}

\noindent{\small\it FIG. 1: $j_0\!=\!2$. Modulus square 
of the amplitude. Left: coherent tetrahedron ($i_{\text{mean}}\!\pm\!\sigma/2=2.54\!\pm\!0.39$). Right: Evolved state ($i_{\text{mean}}\!\pm\!\sigma/2=2.54\!\pm\!0.46$).}

\vspace{0.5cm}
\centerline{\includegraphics[height=3cm]{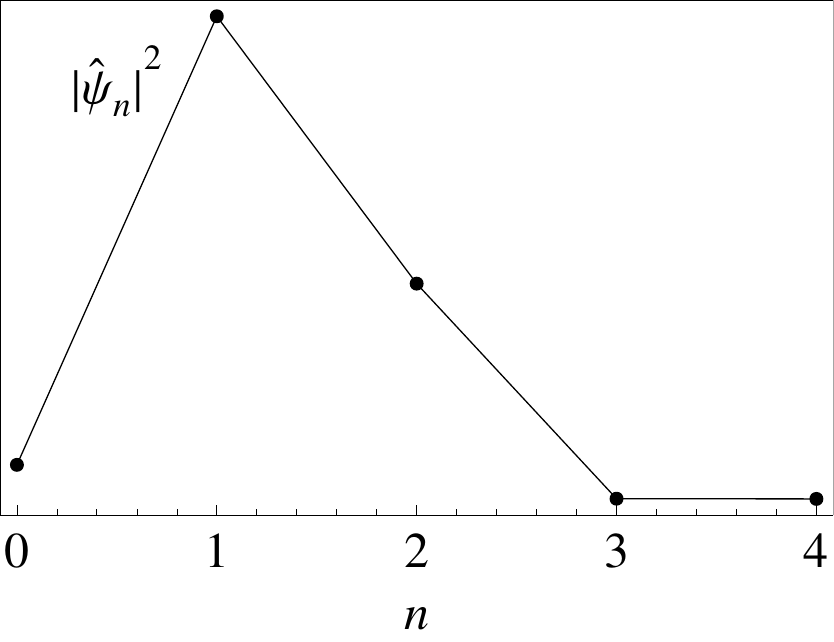}
\hspace{1em} \includegraphics[height=3cm]{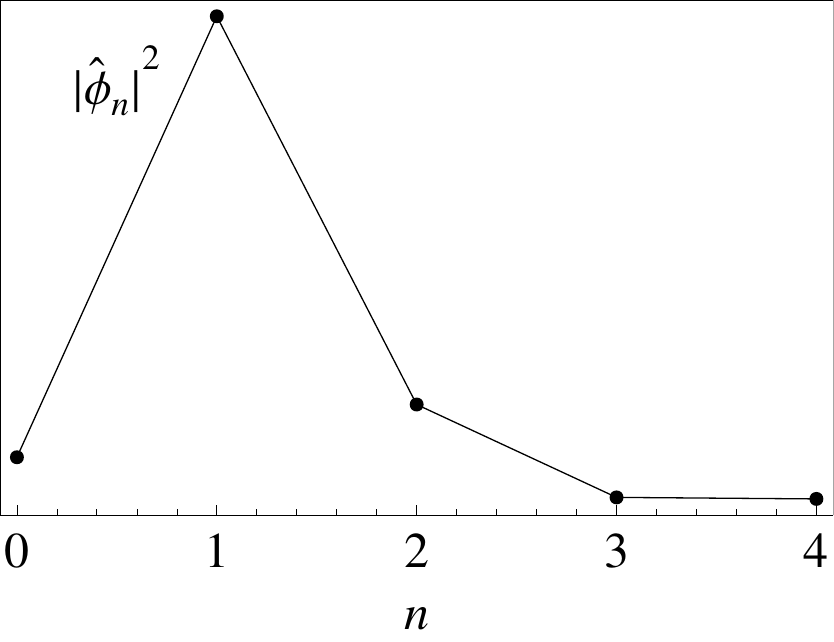}}

\noindent{\small\it  FIG. 2: $j_0\!=\!2$. Modulus square 
of the (discrete) Fourier transform of the amplitude. 
Left: coherent tetrahedron ($n_{\text{mean}}\!\pm\!\sigma/2=1.25\!\pm\!0.27$). 
Right: Evolved state ($n_{\text{mean}}\!\pm\!\sigma/2=1.15\!\pm\!0.31$).}\\

The agreement between the evolved state and the coherent tetrahedron state
is quite good.  Besides the overall shape of the state, notice the concordance 
of the mean values and the widths of the wave packet.  Considering 
the small value of $j_0$, which is far from the large scale limit, and the 
$\tau\to\infty$ limit we have taken, we find this quite surprising. 

The same pattern repeats in the $j_0=4$ case:\\[2mm]

\centerline{\includegraphics[height=3cm]{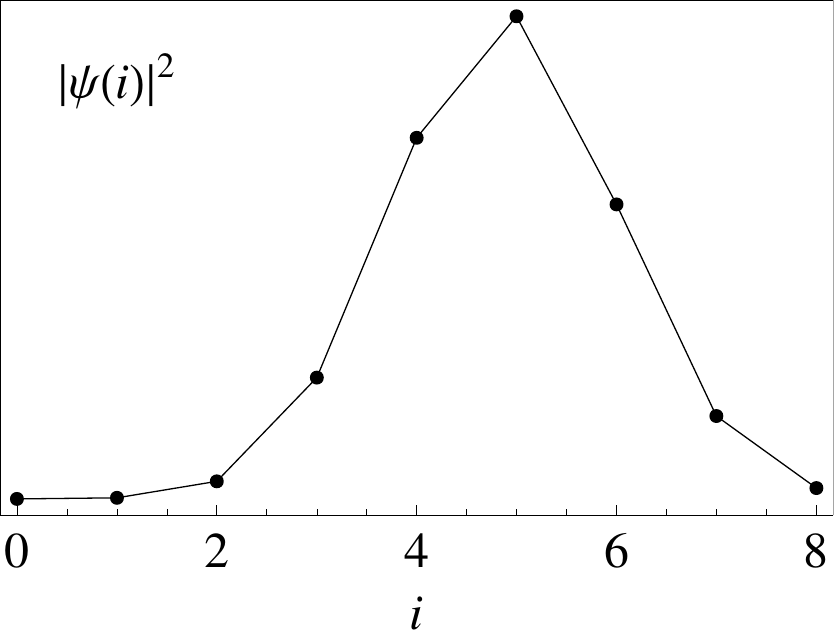}
\hspace{1em} \includegraphics[height=3cm]{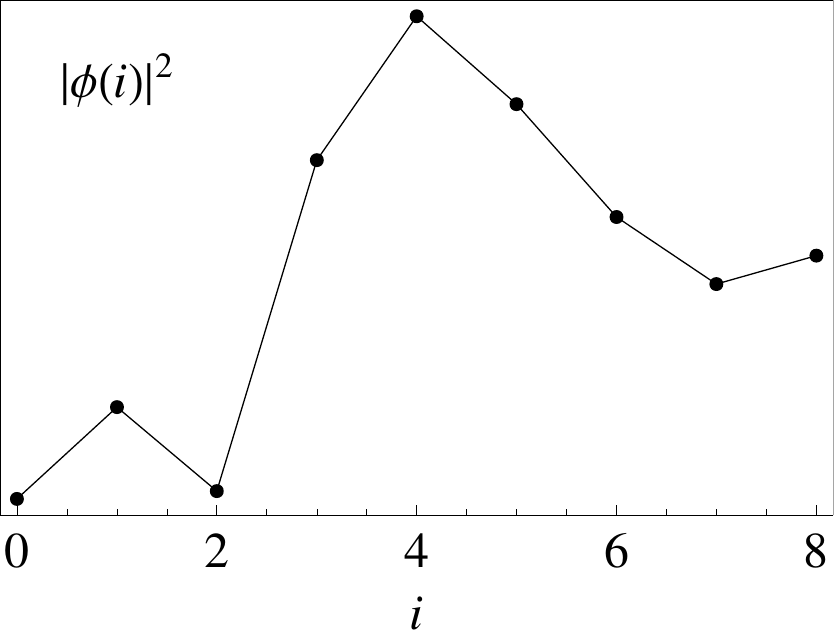}}

\noindent{\small\it  FIG. 3: $j_0\!=\!4$. Modulus square 
of the amplitude. Left: coherent tetrahedron ($i_{\text{mean}}\!\pm\!\sigma/2=4.88\!\pm\!0.56$). Right: Evolved state ($i_{\text{mean}}\!\pm\!\sigma/2=4.85\!\pm\!0.96$).}

\vspace{0.5cm}
\centerline{\includegraphics[height=3cm]{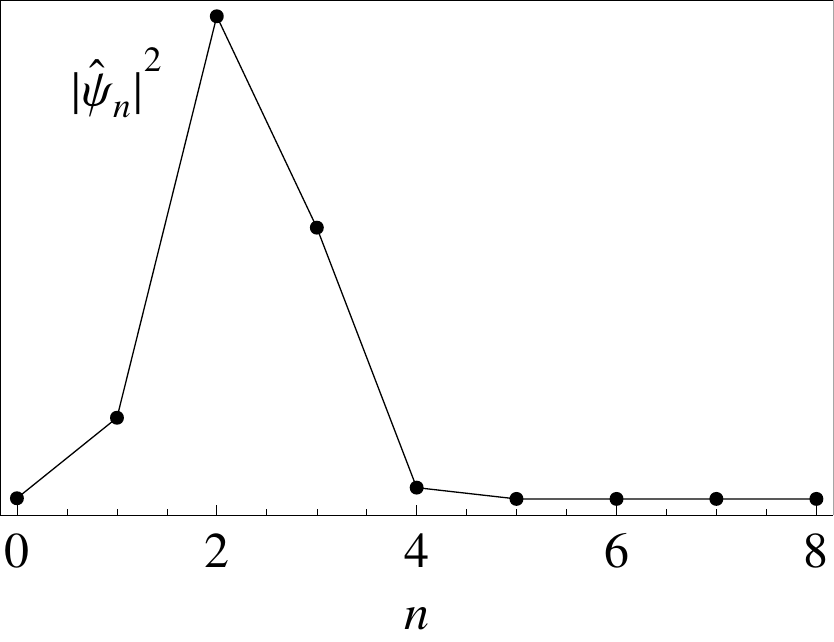}
\hspace{1em} \includegraphics[height=3cm]{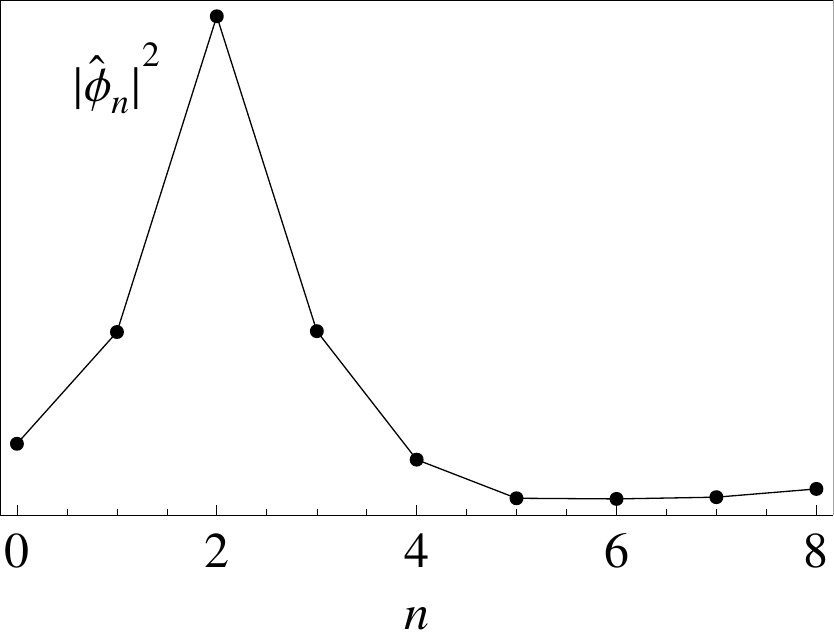}}

\noindent{\small \it FIG. 4: $j_0\!=\!4$. Modulus square 
of the (discrete) Fourier transform of the amplitude. 
Left: coherent tetrahedron ($n_{\text{mean}}\!\pm\!\sigma/2=2.25\!\pm\!0.32$). 
Right: Evolved state ($n_{\text{mean}}\!\pm\!\sigma/2=2.08\!\pm\!0.59$).}

\vspace{0.5cm}

The technique developed here is complementary to the one developed 
in \cite{propagator}, which is  based on computing $n$-point functions. 
We expect that this  technique  could be better developed. 

The numerical results presented above are preliminary and tentative,
but they appear to support the expectation that the flipped vertex might 
in fact give general relativity in the classical limit.\\[3mm]

\noindent We thank Eugenio Bianchi and Etera Livine for numerous useful 
exchanges and suggestions.

%%%%%%%%%%%%%%%%%%%%%%%%%%%%%%%%%%%%%%%%%%%

\end{document}